# Idealizing Tauc Plot for Accurate Bandgap Determination of Semiconductor with UV-Vis: A Case Study for Cubic Boron Arsenide

Hong Zhong[a,†], Fengjiao Pan[b,†], Shuai Yue[c,d,†], Chengzhen Qin[e], Viktor Hadjiev[f], Fei Tian[g], Xinfeng Liu[c,d], Feng Lin[h], Zhiming Wang[i*], Jiming Bao[a,b,e*],

[a]Department of Electrical and Computer Engineering and Texas Center for Superconductivity at the University of Houston (TcSUH), University of Houston, Houston, TX 77204, USA

[b]Department of Physics and Texas Center for Superconductivity at the University of Houston (TcSUH), University of Houston, Houston, TX 77204, USA

[c]CAS Key Laboratory of Standardization and Measurement for Nanotechnology, National Center for Nanoscience and Technology, Beijing 100190, China

[d]University of Chinese Academy of Sciences, Beijing 100049, China

[e]Materials Science & Engineering Program, University of Houston, Houston, Texas 77204, USA

[f]Department of Mechanical Engineering and Texas Center for Superconductivity at the University of Houston (TcSUH), University of Houston, Houston, TX 77204, USA

[g]School of Materials Science and Engineering, Sun Yat-sen University, Guangzhou, Guangdong 510006, China

[h]National Center for International Research on Photoelectric and Energy Materials, School of Materials and Energy, Yunnan University, Kunming 650091, P. R. China

[i]Institute of Fundamental and Frontier Sciences, University of Electronic Science and Technology of China, Chengdu, Sichuan 610054, China

[†]These authors contributed to the work equally

[*]Corresponding authors. Email: zhmwang@uestc.edu.cn, jbao@uh.edu

**Abstract:** The Tauc plot method is widely used to determine the bandgap of semiconductors via UV-visible optical spectroscopy due to its simplicity and perceived accuracy. However, the actual Tauc plot often exhibits significant baseline absorption below the expected bandgap, leading to discrepancies in the calculated bandgap depending on whether the linear fit is extrapolated to zero or non-zero baseline. In this work, we first discuss the origin of background in Tauc plot below the expected bandgap, and show that both extrapolation methods can produce significant errors by simulating Tauc plots with varying levels of baseline absorption. We then propose a new method that involves idealizing the absorption spectrum by removing its baseline before constructing the Tauc plot. By utilizing a gallium phosphide (GaP) wafer with artificially introduced baseline absorptions during UV-Vis measurements, we confirm the detrimental effect of the absorption background and demonstrate the effectiveness of our new method. Finally, we apply this new method to cubic boron arsenide (c-BAs) and resolve discrepancies in c-BAs bandgap values reported by different groups, obtaining a converging bandgap of 1.835 eV based on both previous and new transmission spectra. The method is applicable to both indirect and direct bandgap semiconductors, regardless of whether the absorption spectrum is measured via transmission or diffuse reflectance, will become essential to obtain accurate values of their bandgaps.

**TOC Graphic**

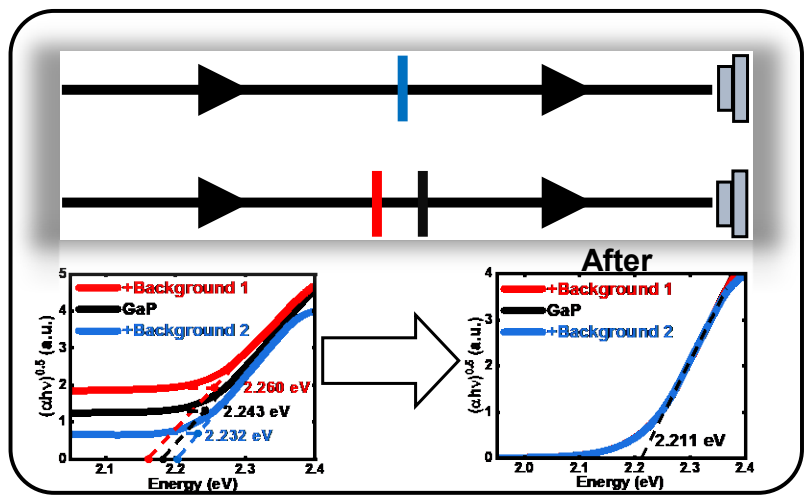

The bandgap is a basic parameter of any semiconductor, and its accurate determination is essential for optimizing their performance in various applications. The most common and accurate method for obtaining the bandgap of thin film or plate semiconductor samples is through the UV-Vis transmission spectrum and the associated Tauc plot[1]. This method is based on extrapolating the linear fit of $(\alpha h\nu)^r$ versus $h\nu$, where $\alpha$ is the absorption spectrum from UV-Vis measurement. The parameter $r$ is 0.5 or 2 depending on whether the semiconductor is an indirect or direct bandgap material, and $h\nu$ is the photon energy [1,2]. Despite its widespread adoption in the materials community, the extrapolation technique remains unsettled and somewhat subjective. This uncertainty arises from non-ideal Tauc plots, such as the one shown in Fig. 1a, which exhibits a persistent non-zero baseline background and does not approach zero at lower energy. Such Tauc plot can lead to two different bandgap values through different extrapolations: a lower bandgap obtained by extrapolating the linear fit to zero of $(\alpha h\nu)^r$, and a higher bandgap obtained through extrapolation to the non-zero baseline. Both extrapolations are widely used in the literature [3-8], and they sometimes even appear in a single figure [6]. Additionally, the definition of the baseline is often not well-defined, with some researchers choosing a flat line from the lowest value in the Tauc plot [4], while others opt for a linear fit to the low-energy portion of the Tauc plot [6].

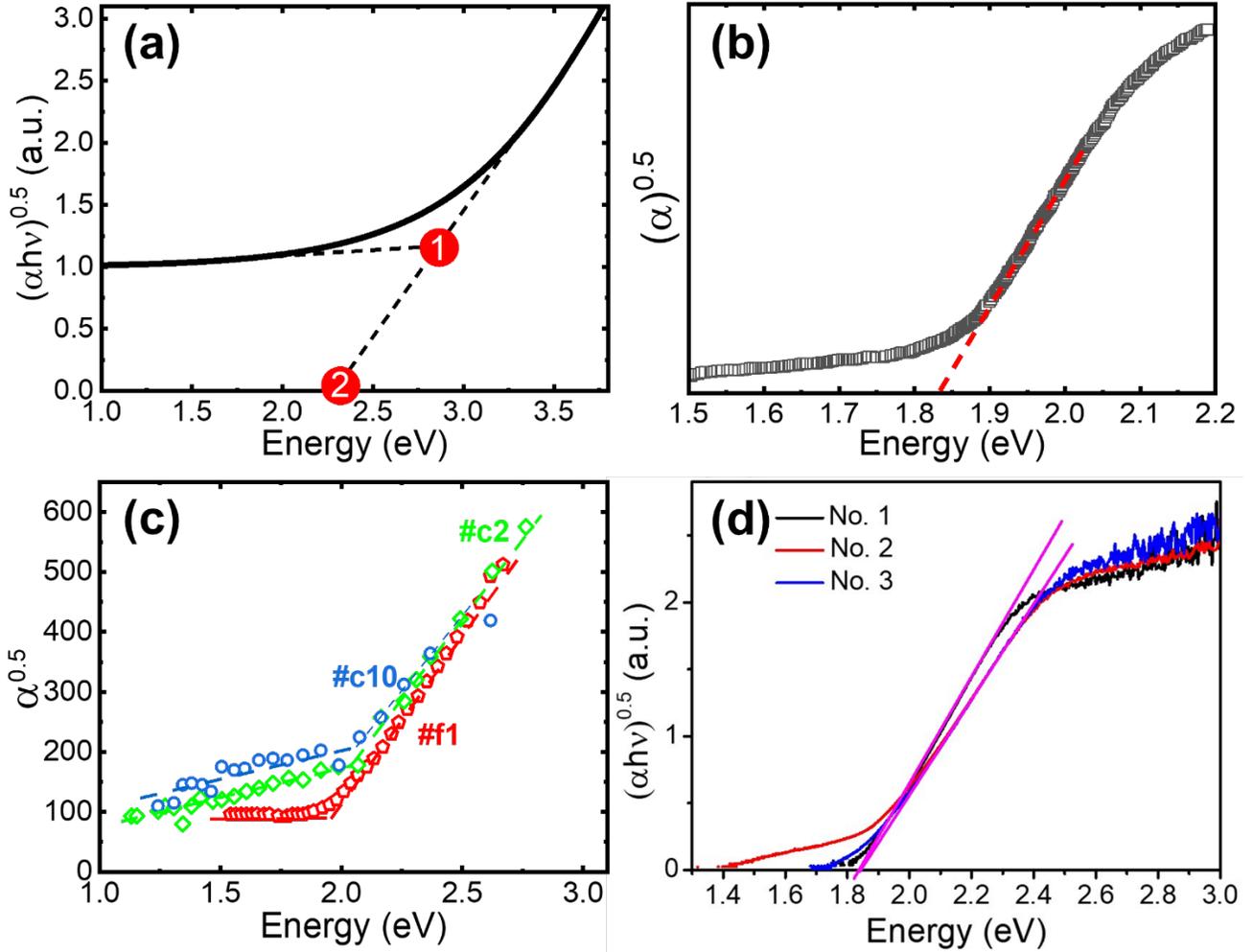

**Figure 1.** Different extrapolations to obtain bandgap from a Tauc plot. (a) Two common extrapolations. (b-d) Three different extrapolations to obtain the bandgap of c-BAs. (b) Reproduced with permission [9], Copyright 2019, AIP Publishing LLC. (c) Reproduced with permission [10], Copyright 2020, Elsevier Ltd. (d) Reproduced with permission [11], Copyright 2020, AIP Publishing LLC.

These different extrapolation techniques have also been used to determine the bandgap of cubic boron arsenide c-BAs, which is a unique semiconductor with simultaneous high thermal conductivity three times that of copper and high carrier mobility three times that of silicon [12-16]. Fig. 1b shows a direct extrapolation to zero and producing a bandgap of 1.82 eV [9], and Fig. 1c shows an extrapolation to the baseline and gives a bandgap of 2.02 eV [10]. It is worth noting that both methods used $\alpha$ instead of $\alpha h\nu$ to obtain the indirect bandgap of c-BAs. The absorption coefficient $\alpha$ is $\sim h\nu\varepsilon_i(h\nu)$, where $\varepsilon_i(h\nu)$ is the imaginary part of dielectric function, which for the indirect bandgap semiconductors is $\varepsilon_i(h\nu)\sim(h\nu)^{-2}(h\nu - E_g \pm E_{ph})^2$ [1,17]. Therefore, $(\alpha h\nu)^{0.5}$, rather than $\alpha^{0.5}$, is expected to be linearized with respect to $h\nu$ [1,2,17]. Additionally, previous work by this group (Fig. 1d) subtracted a constant background absorption to reset $\alpha$ and obtain a Tauc plot [11]. These different measurements and extrapolations have led to a lack of consensus on the bandgap of c-BAs, with an uncertainty of nearly 0.2 eV, a significant deviation from perceived accuracy of UV-Vis measurements. The accurate determination of the bandgap of c-BAs is crucial for its further fundamental study and device applications.

In this work, we first discuss the origin of background in Tauc plot below the expected bandgap, we then propose a model to eliminate the background and obtain an accurate bandgap. By introducing artificial Tauc plot background to a gallium phosphide (GaP) test wafer through numerical modelling and experiments, we investigate and reveal the effect of background on the bandgap. Applying our new model to both GaP and c-BAs, we have not only resolved the discrepancy of bandgap of c-BAs, but also accurately obtained converging values of their bandgaps. Our model and procedure can also be applied to direct bandgap semiconductors and diffuse reflectance. In particular, we will point out some problems with preferred baseline extrapolation in diffuse reflectance Tauc plot [2].

A high-quality single crystal semiconductor should exhibit negligible optical absorption below its bandgap. Although defects may induce sub-bandgap states, the absorption should approach zero toward low energy in an ideal Tauc function $(\alpha h\nu)^r$. However, many Tauc plots do not exhibit this ideal behavior due to several reasons. One major reason is that in typical UV-Vis (or UV-Vis-NIR) measurements, reflectance is often neglected even though it can reduce the incident light intensity by up to 30%. For example, the absorption coefficients of c-BAs in Fig. 1b-c appears to have been used directly without accounting for the reflectance. The second major reason is that surface scratches or dust, grain boundaries, and crystal imperfections can cause scattering of incident light, which can also result in absorption-like optical loss. Other factors such as imperfect beam alignments or measurement can also contribute to non-ideal Tauc plots. Although these non-bandgap absorptions can be strong, they are weakly dependent on wavelength and can be identified and removed from the actual bandgap absorption. Based on these understandings, we propose subtracting the sub-bandgap baseline to idealize the absorption spectrum. This approach will align the theory and experiment of Tauc plot, resulting in a more accurate bandgap value.

Before we apply this new technique to c-BAs, we want to show the effect of baseline on the bandgap determination. Since c-BAs is an indirect bandgap semiconductor, we used $(\alpha h v)^{0.5}$ to calculate Tauc function. We introduced various levels of background absorptions to $\alpha$, mimicking typical experimental scenarios. Fig. 2a displays the ideal Tauc plot with zero absorption below $E_g$, as well as Tauc plots with additional constant background absorptions. It is evident that the background not only introduces a sub-bandgap baseline, but also slows down its transition to linear fit. It is clear that when using traditional extrapolation techniques, the direct extrapolation method significantly underestimates the bandgap, while the baseline extrapolation overestimates it. These results clearly demonstrate that traditional extrapolation techniques are not reliable for accurate bandgap estimation.

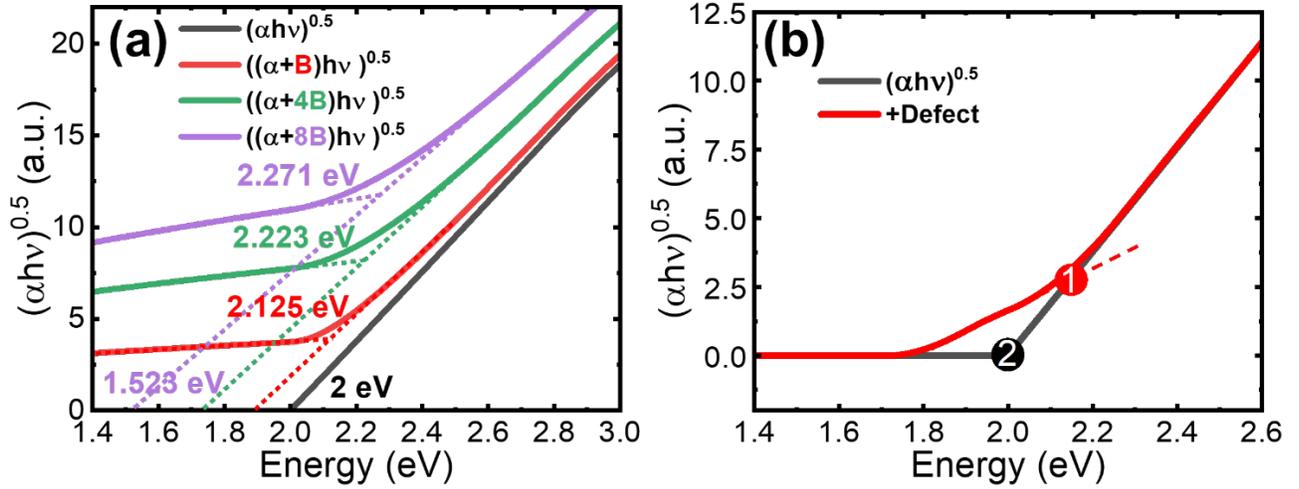

**Figure 2.** Effect of non-ideal Tauc plot on the determination of bandgap through traditional extrapolations. (a) The effect of extended sub-bandgap absorption. (b) The effect of localized defect absorption around the bandgap.

After removing the baseline, some residue absorption may still exist around the bandgap due to defect states. To determine the impact of this absorption on the bandgap calculation, we artificially introduce narrow-band weak absorption and observe its effects. Fig. 2b indicates that while defect absorption may affect the transition from the baseline to linear fit, it does not alter the baseline far below the bandgap and the linear fit far above the bandgap, because its contribution disappears in these regions. Therefore, the bandgap determination using direct extrapolation should not be affected by defect absorption, in contrast to the impact of non-zero baseline on the bandgap, as shown in Fig. 2a.

To verify the results from above simulations, we conducted experiments using a GaP wafer and modified its transmission by inserting different optical components in the sample beam or reference beam. As shown in Fig. 3a, inserting transparent materials in front of the GaP wafer increased the absorption baseline due to the reduced transmission, while adding transparent materials in the reference beam of the UV-Vis decreased the baseline. It is clear that there is persistent absorption below the bandgap even with GaP only. As expected, the resulting Tauc plots of GaP in Fig. 3b also show strong baseline below the bandgap, similar to c-BAs in Fig. 1b-c. Using traditional extrapolation techniques, we observed that an increase in the baseline led to an increase in the bandgap from the baseline extrapolation, while the bandgap from direct extrapolation decreased, in excellent agreement with the numerical modelling in Fig. 2. Therefore, the Tauc plot of the original GaP wafer alone still cannot provide an accurate value of the bandgap when using traditional extrapolation techniques.

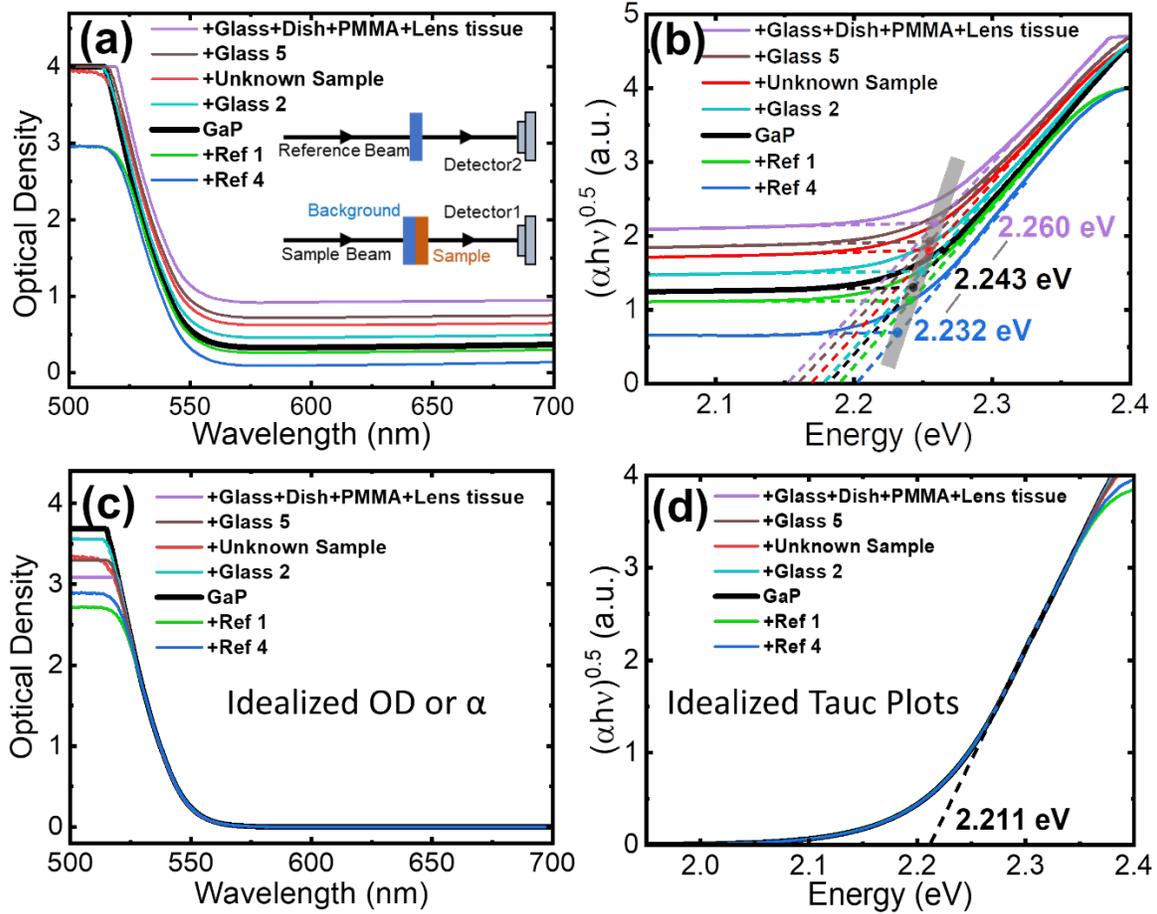

**Figure 3.** Measured absorptions, Tauc plots and corresponding bandgaps of a GaP wafer with different optical materials. Inset of (a): Method to change the measured absorption spectrum of GaP. (a) Before and (c) after the application of idealization of absorptions. (b, d) Corresponding Tauc plots of (a) and (c) respectively.

Despite these seemingly different bandgap values, there should be only one actual value since they all come from the same GaP wafer. Now we can try the proposed idea to obtain a more reliable bandgap value. The first step is to clean up the absorption spectrum by removing the background, which can be linearly fit. Fig. 3c shows that after subtracting the baseline, the absorption spectra from different measurements have zero baseline absorption and overlap with each other. As a result, the corresponding Tauc plots in Fig. 3d also overlap perfectly, producing the same bandgap value of 2.21 eV, in excellent agreement with earlier work [18].

With strong theoretical understanding and experimental verification, we are confident in applying this method to c-BAs. We started with Fig. 1c, #f1, which initially gave a high bandgap of ~2.0 eV. After removing the background absorption, Fig. 4a provided a new bandgap of 1.834 eV. Similarly, for Fig. 1b, the bandgap increased from 1.824 to 1.838 eV (Fig. 4b) after background removal. Here we still use the same $\alpha^{0.5}$ to obtain the bandgap for a better comparison. To further verify our method, we prepared three new c-BAs samples, as shown in Fig. 4c, all of which had surface scratches. Fig. 4d shows their Tauc plots obtained directly from UV-Vis in Fig. 4c, with two samples exhibiting increased absorption at longer wavelengths. Fig. 4e shows their idealized absorption spectra after removing the long wavelength background, and the corresponding Tauc plots are shown in Fig. 4f. Surprisingly, extrapolations of all three samples pointed to the same bandgap of 1.836 eV. Based on the results from both old and new data, we believe that the bandgap of c-BAs should be $1.835 \pm 0.005$ eV.

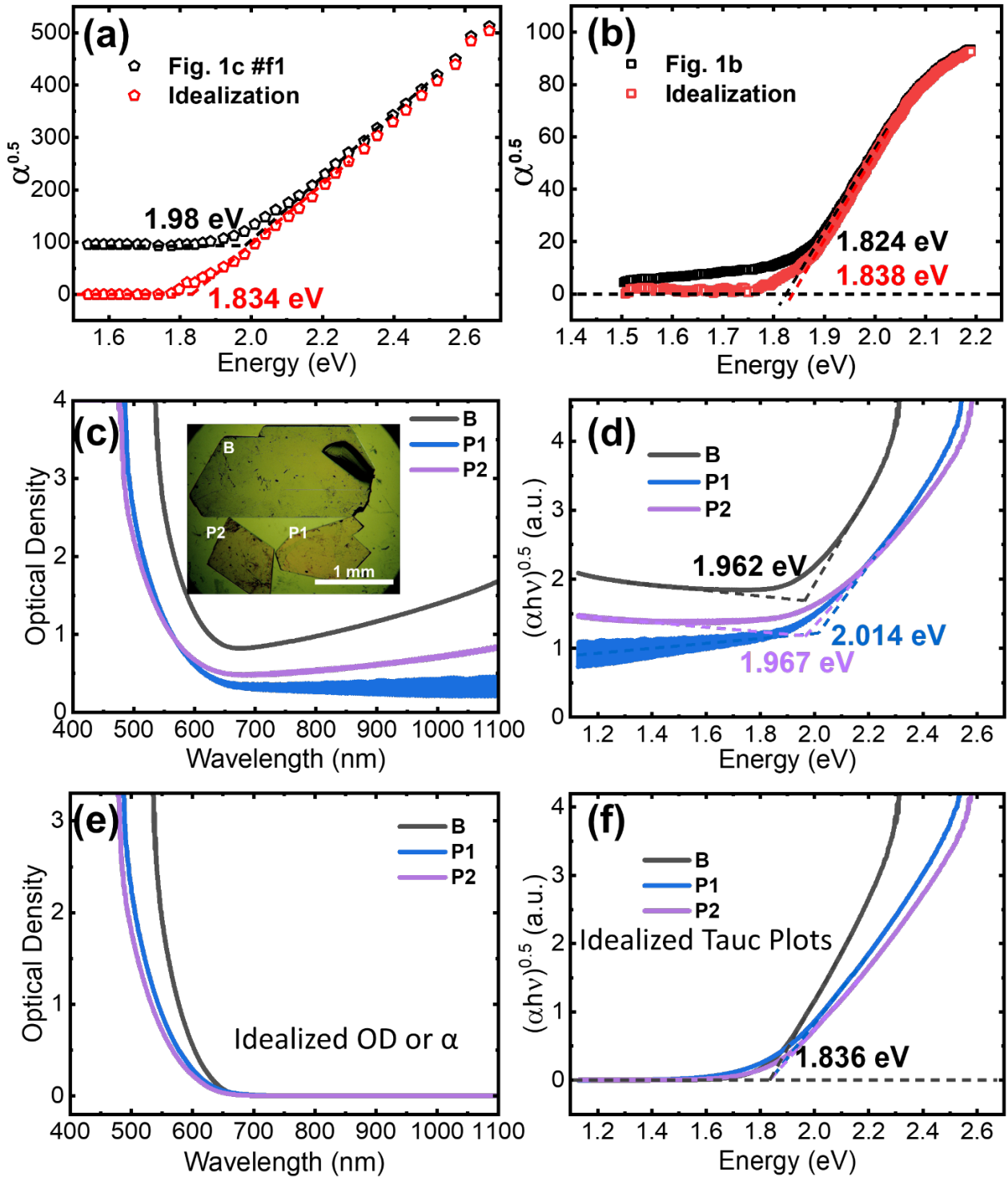

**Figure 4.** Bandgap of c-BAs using new method. (a-b) Idealized Tauc plots and new bandgaps from (a) Fig. 1c and (b) Fig. 1b. (c) Pictures of three new samples and their UV-Vis spectra. (d) Tauc plots and higher bandgaps based on (c). (e) Revised absorption spectra of (c) after subtraction of background. (f) Corresponding Tauc plots of (e) showing the same bandgap.

Compared to the reported bandgap of 2.02 eV [10,15], the revised bandgap of 1.83 eV for c-BAs is more consistent with its photoluminescence (PL) at room temperature. The PL peak energy ($E_{PL}$) is determined by the formula $E_{PL} = E_g - E_b - E_{ph}$, where $E_g$ is the bandgap at room temperature, $E_b$ is the binding energy of electron and hole pair (exciton), and $E_{ph}$ is the phonon energy [17]. The $E_b$ was estimated to be ~ 40 meV [19,20], while $E_{ph}$ varies from 20 to 40 meV at the X point and 80 meV at the Γ point [21]. Based on our own works [11,16,22], the room temperature PL peak at 720 nm (1.72 eV) indicates that the bandgap $E_g$ should fall in the range of 1.78 – 1.84 eV. Even with a shorter wavelength PL peak of 695 nm (1.78 eV) reported in [15], the bandgap still falls within the range of 1.84 – 1.90 eV. Both ranges are still below the reported value of 2.02 eV, but closer to 1.84 eV. It should be noted that the reported value of 2.02 eV was supported by a DFT calculation [10,20,23]; however, other DFT studies produced a lower bandgap of 1.78 eV at zero temperature [19]. Furthermore, according to Bravic *et al.*, the bandgap will decrease with increasing temperature, and they predicted an even lower bandgap of 1.52 eV at zero temperature [24].

Besides UV-Vis transmission spectroscopy, diffuse reflectance is also a widely used technique for measuring the optical absorption spectrum of powders or particles. To obtain their bandgap through Tauc plot, the KM function *F(R)* is calculated based on diffuse reflectance *R*, and is then used as an equivalent absorption spectrum *α*. Like transmission spectroscopy, Tauc plots based on diffuse reflectance also exhibit features such as a strong sub-bandgap absorption baseline [25-27], no baseline [25-27], or strong baseline absorption near the expected bandgap [2,28]. We believe that a strong baseline in *F(R)*-based Tauc plot will also lead to errors similar to those in transmission-based Tauc plot. Thus, it is recommended that Tauc plots be idealized through *F(R)* before performing extrapolation, due to the equivalence between *F(R)* and *α*. In cases where there is no low energy background but strong absorption near the bandgap, it is important to understand the nature of the absorption before extrapolation. In the case of a mixture of $TiO_2$ and dye, it is reasonable to use the dye absorption as a baseline for extrapolation, since the dye has an extended strong absorption beyond the bandgap of $TiO_2$, which is what linear extrapolation assumes [2]. In other cases where only a single homogeneous material is involved and absorption occurs only near the bandgap as shown in Fig. 2b, the baseline can be neglected and direct extrapolation should be performed to obtain a more accurate bandgap than baseline extrapolation [2].

In summary, our findings demonstrate that conventional extrapolation methods in Tauc plots from UV-Vis transmission or diffuse reflectance may lead to inaccurate estimation of the bandgap of semiconductors. To obtain a more precise value, it is crucial to subtract background absorption from the original absorption spectrum prior to making an ideal Tauc plot and performing extrapolation to determine a unique and accurate bandgap. This new approach has been verified through numerical modelling and experimental validation using GaP as a reference material. By applying this method, we have resolved the bandgap discrepancy of c-BAs and obtained a more precise value of 1.835 eV. As new materials continue to emerge and Tauc plot remains a widely used technique, our proposed method provides the materials community with a more reliable tool to investigate fundamental properties and explore potential applications [28-35].

**Notes**

The authors declare no competing financial interest.